\documentclass[%
prl,
reprint
]{revtex4-1}
\usepackage{graphicx}
\usepackage[update,prepend]{epstopdf}

\usepackage{xcolor}

\newcommand{\LN}{LiNbO\textsubscript{3}}

\usepackage{siunitx}
\usepackage{amsmath}
\usepackage{mathtools}

\DeclareSIUnit{\belmilliwatt}{Bm}
\DeclareSIUnit{\dBm}{\deci\belmilliwatt}

\begin{document}

\title{An ultra-stable microresonator-based electro-optic dual frequency comb}

\author{N. J. Lambert}
\email{nicholas.lambert@otago.ac.nz}
\affiliation{Department of Physics, University of Otago, Dunedin, New Zealand}
\affiliation{The Dodd-Walls Centre for Photonic and Quantum Technologies, New Zealand}
\author{L. S. Trainor}
\affiliation{Department of Physics, University of Otago, Dunedin, New Zealand}
\affiliation{The Dodd-Walls Centre for Photonic and Quantum Technologies, New Zealand}
\author{H. G. L. Schwefel}
\affiliation{Department of Physics, University of Otago, Dunedin, New Zealand}
\affiliation{The Dodd-Walls Centre for Photonic and Quantum Technologies, New Zealand}

\date{\today}

\maketitle

\textbf{Optical frequency combs emit narrow pulses of light with a stable repetition rate. Equivalently, the generated light spectrum consists of many discrete frequencies spaced by this same repetition rate. These precision light sources have become ubiquitous in applications of photonic technologies~\cite{diddamsOpticalFrequencyCombs2020, pfeifle_coherent_2014, ataie_ultrahigh_2015, pappMicroresonatorFrequencyComb2014, obrzudMicrophotonicAstrocomb2019, lucasUltralownoisePhotonicMicrowave2020} because they allow coherent sampling over a broad part of the optical spectrum~\cite{millotFrequencyagileDualcombSpectroscopy2016}. The addition of another comb, with a slightly different line spacing, results in a \emph{dual comb}. Widely used in spectroscopy~\cite{coddingtonCoherentDualcombSpectroscopy2010,suhMicroresonatorSolitonDualcomb2016,ycasHighcoherenceMidinfraredDualcomb2018}, dual combs allow one to read out the broad frequency response of a sample in a simple electronic measurement~\cite{muravievMassivelyParallelSensing2018}. Many dual comb applications require a high level of mutual coherence between the combs~\cite{nishiyamaDopplerfreeDualcombSpectroscopy2016}, but achieving this stability can be demanding. Here, by exploiting the rich structure of the nonlinear electro-optic tensor in lithium niobate, we generate ultra-stable dual combs with the two combs naturally having orthogonal polarizations. Our combs have relative linewidths down to 400 microhertz, and require no stabilization or post-processing methods. The ultra-high stability of the spectrum emitted by our device, along with its simplicity of operation and energy efficiency, offer a route to the deployment of robust and versatile dual comb sources.}

A dual frequency comb allows one comb to be used as a probe and the other to be used as a reference against which the probe can be compared. For example, by measuring the beat frequencies between close-in-frequency comb lines, spectroscopic analysis can take place in the radiofrequency (r.f.) part of the electromagnetic spectrum, allowing fast and economical measurement. To exploit this in full, a narrow relative linewidth between the comb line pairs~\cite{muravievMassivelyParallelSensing2018, nishiyamaDopplerfreeDualcombSpectroscopy2016} is required. This precise synchronization between the combs permits long coherent averaging periods, improving sensitivity and increasing signal-to-noise ratios. For instance, the distance uncertainty for dual comb based distance measurements~\cite{coddingtonRapidPreciseAbsolute2009, suhSolitonMicrocombRange2018} is proportional to the optical phase uncertainty of the comb lines~\cite{trochaUltrafastOpticalRanging2018}. Tight locking of two pulsed lasers provides a stable spectrum~\cite{coddingtonCoherentMultiheterodyneSpectroscopy2008,coddingtonCoherentDualcombSpectroscopy2010}, but comes with high complexity and cost, making operation outside of a laboratory environment difficult. Analogue~\cite{ideguchiAdaptiveRealtimeDualcomb2014} and digital~\cite{royContinuousRealtimeCorrection2012} correction techniques have also been applied, but increase the data handling requirements. The generation of stable dual combs therefore remains an important challenge.

Here we present a straightforward method of generating electro-optic dual combs with extremely high mutual coherence from a single continuous-wave laser source. It makes use of the optical mode structure in microresonators to generate two combs of orthogonal polarization, allowing them to be easily separated into separate beam paths. Our device is compact, and -- because it is resonant to all of the involved electromagnetic fields -- it is efficient. The resulting dual comb line pairs have relative stabilities as low as $\sim\SI{400}{\micro\hertz}$, and our method is fully free-running, requiring no complicated stabilization techniques or digital post-processing. Moreover, there are no thermal instabilities to navigate during comb formation~\cite{leshemThermalInstabilitiesFrequencycomb2021}, nor is there any need for careful thermal control~\cite{joshiThermallyControlledComb2016}.

Our device exploits whispering gallery modes (WGMs) supported by a nonlinear optical microresonator~\cite{grudininUltraHighCrystalline2006,reviewdmitry}. The even spacing and high $Q$ of these modes makes them a natural platform for efficient frequency comb generation, and the nonlinearity allows the generation of new frequencies. Nonlinearities due to the third-order susceptibility $\chi^{(3)}$ have been used extensively to produce solitonic behaviour in the propagating light field, allowing the generation of Kerr-soliton frequency combs~\cite{delhayeOpticalFrequencyComb2007}. 
Dual Kerr combs have been demonstrated~\cite{suhMicroresonatorSolitonDualcomb2016}, including separable counter-propagating dual combs~\cite{yangCounterpropagatingSolitonsMicroresonators2017, lucasSpatialMultiplexingSoliton2018}, but require sophisticated techniques for stabilization \cite{kippenbergDissipativeKerrSolitons2018}.

Here, we instead use a second order, or electro-optic, nonlinearity for comb generation~\cite{behaElectronicSynthesisLight2017,kovacichShortpulsePropertiesOptical2000}, via cascaded sum- and difference-frequency generation between the optical carrier and an incident microwave drive tone. Because the comb line spacing is given by the microwave frequency, this technique allows more straightforward control of comb repetition rates. By applying two microwave tones simultaneously, dual~\cite{millotFrequencyagileDualcombSpectroscopy2016, duranElectroopticDualcombInterferometry2016} and dual-driven combs can be generated~\cite{zhangBroadbandElectroopticFrequency2019}.

\begin{figure*}
\includegraphics{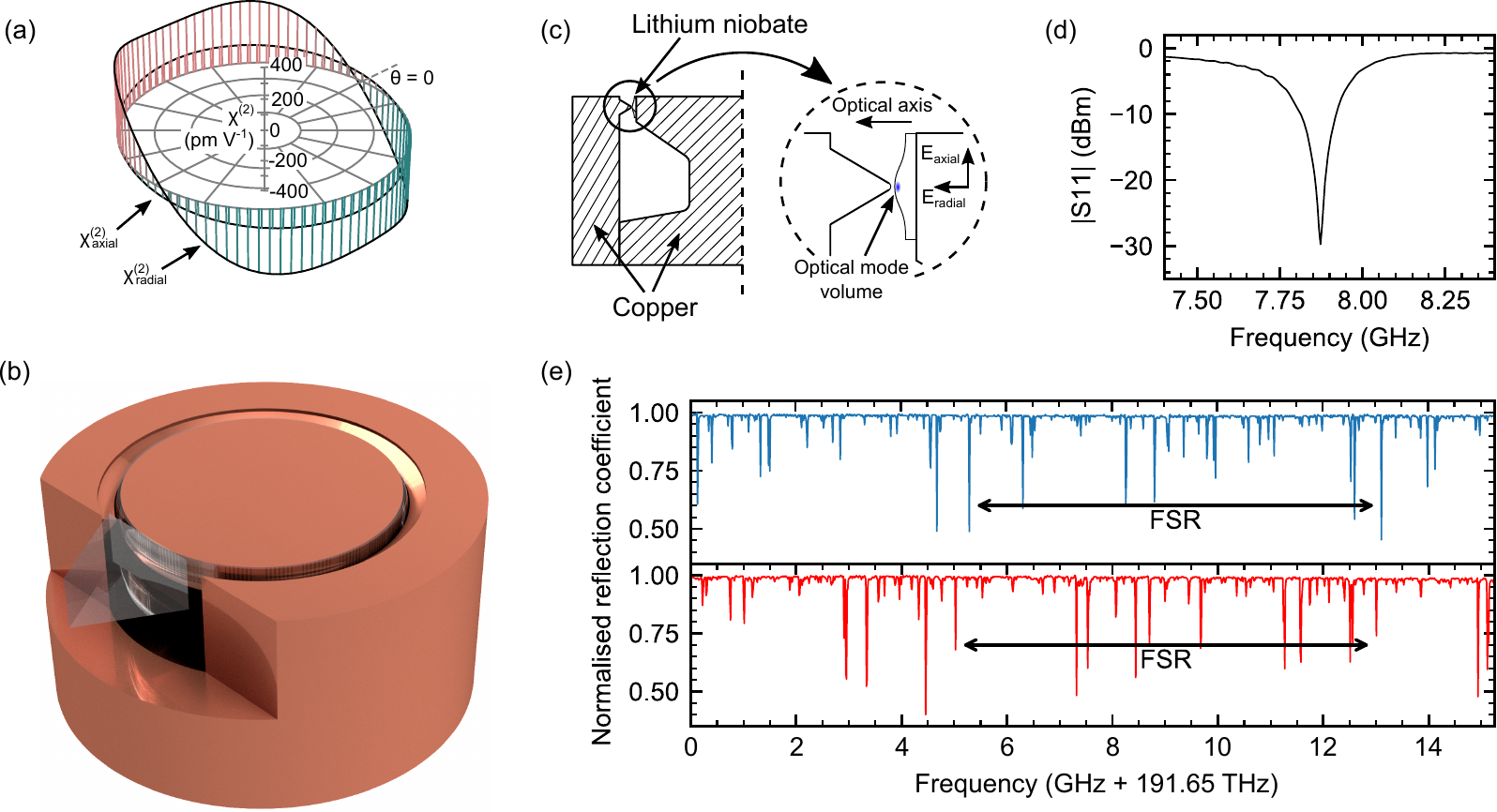}
\caption{\textbf{Device principles and design.} (a) Effective electro-optic coefficient of \LN{} for axial and radial modes as a function of angle from the crystallographic $z$-axis, showing the angular dependency. (b) Cartoon of our device, with the \LN{} ring embedded in a toroidal loop gap. A gap in the outer of the cavity provides for optical access via prism coupling. (c) Left - half cross-section of the device in (b). Right - expanded view of \LN{} resonator, in the gap of the loop-gap resonator, showing the location of the optical mode volume.  (d) Microwave spectrum of the cavity, tuned to critical coupling. (e) Optical mode spectrum for radial (upper) and axial (lower) modes. FSRs are arrowed.} \label{fig:phase_match_scheme}
\end{figure*}

Figure \ref{fig:phase_match_scheme}(b) shows a schematic illustration of our implementation. At the core of our device is a ring-shaped microresonator made from mechanically polished $x$-cut lithium niobate (\LN{})~\cite{PhysRevApplied.9.024007, sedlmeirExperimentalCharacterizationUniaxial2013}. The WGMs ($Q\sim10^8$) are driven with CW laser light at a wavelength close to \SI{1550}{\nano\metre}. The resonator can support two optical mode families with different free spectral ranges; axial modes, in which the electric field is normal to the plane of the resonator, and radial modes, for which the electric fields lies in the plane of the disc. Unlike \LN{} resonators with the $z$-axis normal to the plane of the WGMs, an $x$-cut resonator has a point on its rim for which the wavevector of light of both polarizations is equal. This allows efficient evanescent coupling to both axial and radial modes using light with the same angle of incidence.

To permit the generation of new optical frequencies, the optical microresonator is embedded in a toroidal loop gap microwave cavity (Fig.~\ref{fig:phase_match_scheme}(b,c), see Methods and Supplementary Information). The electric field vector of the lowest order microwave mode is constant in magnitude around the torus, and points between the inner and outer of the cavity. It is therefore aligned with the radial axis of the microresonator. By fashioning a sharp edge on the outer surface of the loop capacitance, we focus the electric field into the optical mode volume, increasing the overlap between optical and microwave fields. The microwave cavity mode has a centre frequency of \SI{7.87}{\giga\hertz}, lying between the free spectral ranges (FSRs) of the radial and axial mode families. By driving the cavity at one of these FSRs using a coupled antenna, a frequency comb can be generated; by driving it simultaneously at both FSRs, a dual comb results.

The interaction of different frequencies in nonlinear materials generally requires careful consideration of phase-matching, in order to preserve both energy and momentum~\cite{reviewdmitry}. This requirement is encapsulated by the expression for the coupling rate $g$ between the complex electric fields of the input optical and microwave modes, $E_\textrm{in}$ and $E_\Omega$ respectively, and that of the mode $E_\textrm{out}$ at the generated optical frequency,
\begin{equation}
g \propto \int\displaylimits_{\mathclap{\substack{\textrm{mode} \\ \textrm{volume}}}}    \chi^{(2)}E^{}_\textrm{in}E^*_\textrm{out}E^{}_\Omega dV.
\end{equation}
Here, the mode volume integral runs around the resonator. Because the input and output modes are orthogonal, spatially uniform microwave modes and electro-optic coefficients lead to $g=0$. Therefore, for efficient comb generation, either the microwave mode or the electro-optic coefficient must have an antisymmetric spatial component.

In previous work using $z$-cut \LN{}, the microwave field was engineered to have a large antisymmetric component, for example by using an electrode on only one side of the WGM resonator~\cite{ilchenko_whispering-gallery-mode_2003} or by driving higher order modes of a microwave cavity~\cite{Rueda19, Rueda:16}. In $x$-cut \LN{}, an alternative approach is possible. Here, the effective nonlinearity for light confined to the rim of the resonator varies azimuthally around the resonator's rim, because the effective $\chi^{(2)}$ is dependent on the direction of the wavevector of the light; the light propagating around the resonator therefore experiences an oscillating $\chi^{(2)}$ (Fig \ref{fig:phase_match_scheme}(a)). This spatial variation allows a uniform microwave field to couple two different spatially orthogonal optical modes from the same mode family. The effective nonlinearities for the modes, $\chi^{(2)}_\textrm{eff}$, are given by the Fourier component of $\chi^{(2)}(\theta)$, which provides the necessary number of momentum quanta (see Supplementary Information). For the axial modes we find $\chi^{(2)}_\textrm{eff}=\SI{63.2}{\pico\metre\per\volt}$ and for radial modes $\chi^{(2)}_\textrm{eff}=\SI{240}{\pico\metre\per\volt}$ for \SI{633}{\nano\metre} light and \SIrange{50}{86}{\mega\hertz} modulation frequency, comparable to the largest component of the electro-optic tensor $\chi^{(2)}_{33} = \SI{362}{\pico\metre\per\volt}$~\cite{turnerHIGHFREQUENCYELECTRO1966}, as used in $z$-cut devices.

\begin{figure}[t!]
\includegraphics{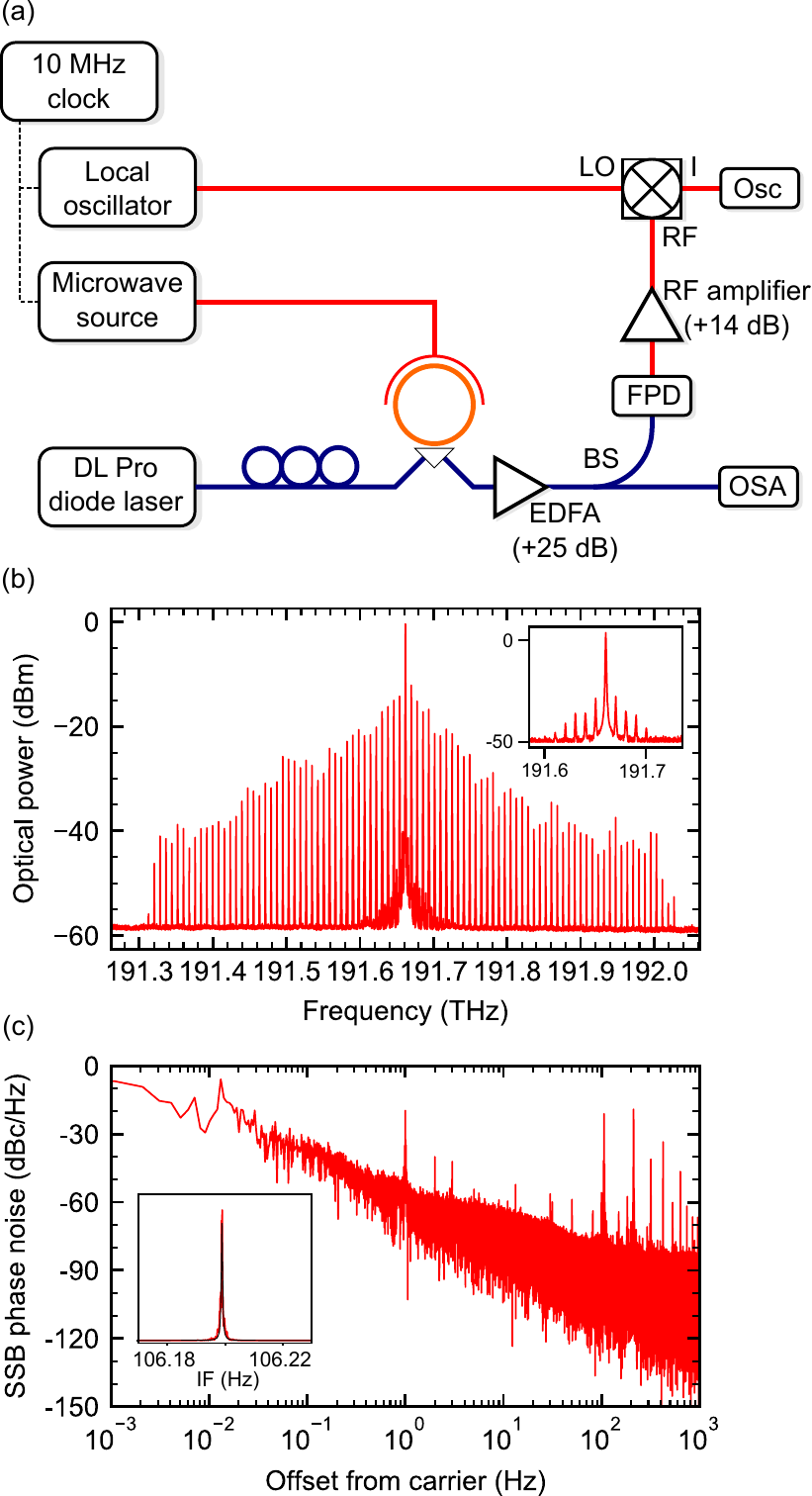}
\caption{\textbf{Single comb generation.} (a) Experimental set up. The output of a grating stabilized diode laser is tuned to a mode of a thermally stabilized WGM resonator. The surrounding microwave cavity is driven at the FSR of the optical modes, and the resulting comb is amplified with an erbium-doped fibre amplifier (EDFA) before being measured with both an optical spectrum analyser (OSA) and a fast photodiode (FPD). The FPD output is mixed with a local oscillator detuned by $\sim\SI{100}{\hertz}$ from the comb repetition rate, and the resulting IF signal digitized. (b) A single frequency comb, with 91 comb lines visible and a repetition rate of \SI{7.940}{\giga\hertz}. Inset is the structure of the laser line, measured with microwaves off. (c) SSB phase noise of the IF due to a single comb, as a function of offset from the carrier. Inset is the IF lineshape (red), and its fit to a Lorentzian lineshape (black).}\label{fig:exp_config}
\end{figure}

We first characterize the single comb that is generated when a microwave field with a single frequency interacts with the optical field. \SI{1550}{\nano\metre} light from a Toptica DL Pro grating stabilized diode laser is passed through a fibre polarization controller and then coupled into the WGM resonator (major radius = \SI{2.56}{\milli\metre}, minor radius = \SI{400}{\micro\metre}) using a GRIN lens and diamond prism (Fig.~\ref{fig:exp_config}(a)). The emitted light from the resonator is out-coupled in the same way. We chose here to use a radially polarized optical mode family, with $Q$ factors approaching $10^8$ and an FSR of \SI{7.940}{\giga\hertz}; we therefore drive the microwave cavity at this frequency, with a incident power on the cavity of \SI{35}{\dBm}. We measure the resulting output spectrum with an optical spectrum analyser and find that a single comb with 91 comb lines is generated (Fig.~\ref{fig:exp_config}(b)). The comb spectrum is cut off abruptly at \SI{191.3}{\tera\hertz} and \SI{192.05}{\tera\hertz}; at these points dispersion-induced breakdown of the comb occurs, as the intrinsic and geometric dispersion of the resonator results in a change of FSR at large detunings from the centre frequency~\cite{Rueda19,zhangBroadbandElectroopticFrequency2019}.

To measure the stability of the repetition rate of the comb, the intensity of the frequency comb is measured with a fast photodiode with a bandwidth of \SI{20}{\giga\hertz}. The output from this carries the beat frequency between comb lines, and the linewidth of this signal is a measure of the comb spacing stability. This linewidth is too small to measure directly using, for example, a microwave spectrum analyser. Instead, we mix this signal with a local oscillator detuned from the microwave source by \SI{106.2}{\hertz}, and study the linewidth of the resulting intermediate frequency (IF) by digitally sampling it and taking the power spectrum of the resulting discrete time signal (see Methods). In Fig.~\ref{fig:exp_config}(c) we show the single sideband (SSB) phase noise of the IF, with its lineshape inset. We find a linewidth of $\sigma_\textrm{meas}=\SI{0.24(7)}{\milli\hertz}$, demonstrating extremely high stability of the repetition rate of the frequency comb. 

Contributions to $\sigma_\textrm{meas}$ come from phase noise on the microwave drive signal and optical phase noise due to the lithium niobate. To quantify these components, we start by directly measuring the linewidth of the microwave drive tone, using the same r.f.~measurement chain. We find it to be $\sigma_\textrm{mW}=\SI{0.100(019)}{\milli\hertz}$. The dominant source of the \SI{7.940}{\giga\hertz} microwave signal from the comb is the beat note between the carrier and the two first order comb lines, which are summed coherently. We therefore deduce that noise processes in the lithium niobate contribute an effective linewidth $\sigma_\textrm{LN}=\sqrt{\sigma_\textrm{meas}^2-\sigma_\textrm{mW}^2}= \SI{0.22(08)}{\milli\hertz}$ for the first order comb lines.

\begin{figure*}
\includegraphics{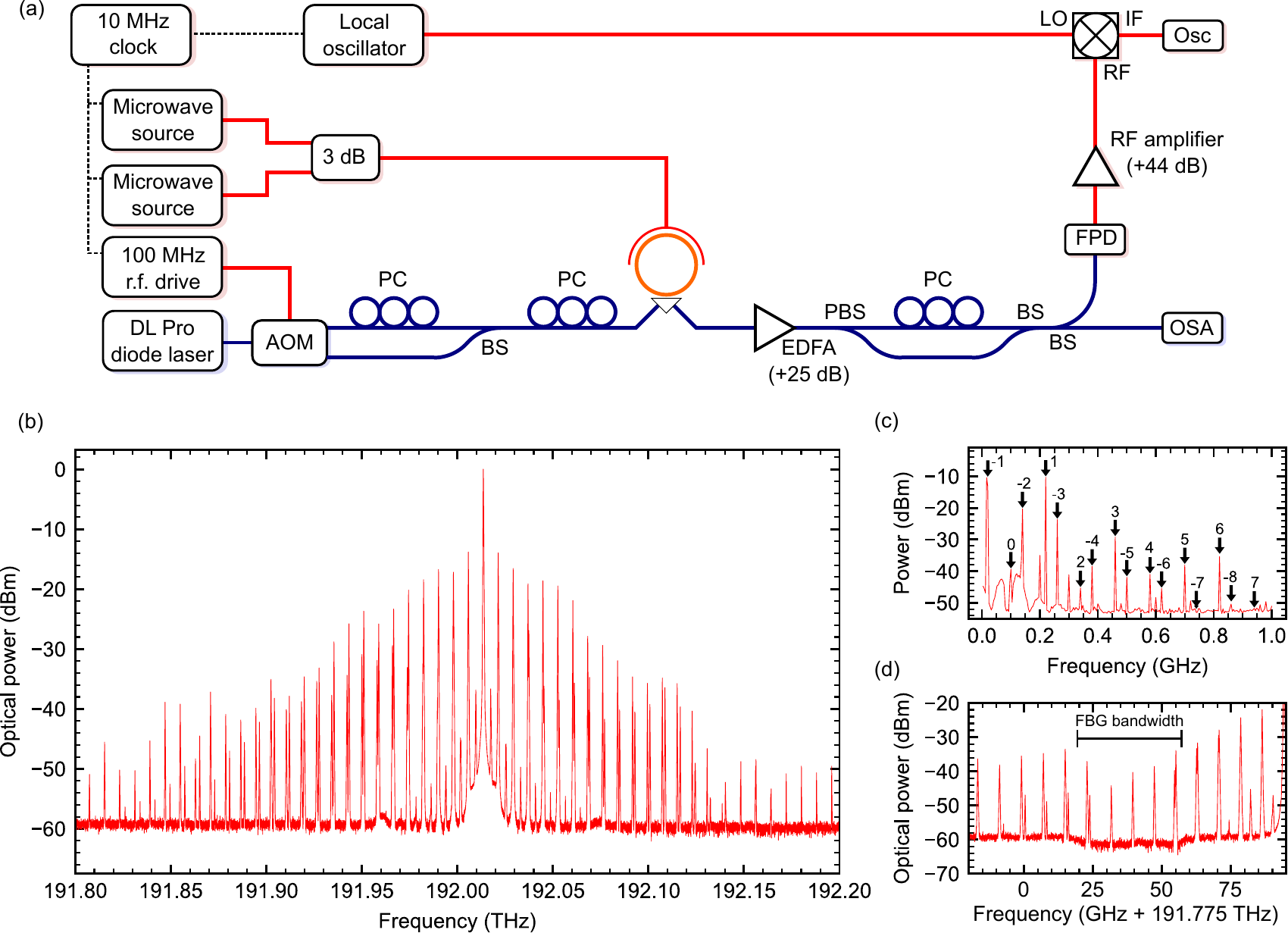}
\caption{\textbf{Dual comb generation.} (a) Experimental set-up for dual comb generation. Two frequencies resonant with modes of orthogonal polarization are generated from a single laser and an AOM. The cavity is simultaneously driven at the FSR of both mode families. The resulting combs are separated using a polarising beam splitter. The reference comb is then rotated on to the same polarization as the probe before output to our measurement chain. (b) A dual polarization dual comb, with line spacings of \SI{7.934}{\giga\hertz} and \SI{7.814}{\giga\hertz}. (c) Mixing of dual comb lines to RF frequencies. Beat frequencies are labelled with the order of the originating comb lines. (d) Proof of principle spectroscopy, showing lines from the `probe' comb at frequencies between \SI{191.787}{\giga\hertz} and \SI{191.827}{\giga\hertz} being filtered by a FBG.}\label{fig:dualcomb}
\end{figure*}

We now exploit the two polarizations supported in the WGM resonator to generate a dual comb. We begin by using an acousto-optic modulator (AOM) to split the carrier into two equal magnitude carriers separated by \SI{100}{\mega\hertz} (Fig.~\ref{fig:dualcomb}(a)). These are prepared in orthogonal linear polarization states, and then recombined into a single fibre. The polarizations are then rotated so that they correspond to axial and radial modes in the WGM resonator, and can therefore excite two optical modes of different polarization with centre frequencies \SI{100}{\mega\hertz} apart.

Two microwave tones at \SI{7.814}{\giga\hertz} and \SI{7.934}{\giga\hertz}, commensurate with the FSRs of the axial and radial modes respectively, are then  excited in the metal cavity. Two frequency combs are observed, with comb line spacings equal to the applied microwave tones (Fig.~\ref{fig:dualcomb}(b)). The longer comb, with more than 50 visible comb lines, forms in the radial mode family, while the axially polarized comb is shorter due to its lower $\chi^{(2)}_\textrm{eff}$.

A key advantage of dual comb techniques is that the combs can traverse spatially separated paths, and then be referenced against each other by mixing them down to r.f.~frequencies with a fast photodiode. This requires that they can be separated into `probe' and `reference' combs; for the output of our device this can be achieved straightforwardly by a polarising beam splitter, as the generated combs are orthogonally polarized. In order to mix them together, the reference comb polarization is then rotated \ang{90} so that it has the same polarization as the probe comb, and they are then recombined with a \SI{3}{\decibel} beam splitter before the photodiode.

In Fig.~\ref{fig:dualcomb}(c) we show the low frequency regime of the resulting spectrum. We label the peaks with the order of the originating comb lines, with $n=0$ labelling the beating of the two pump tones. We also observe artefacts at frequencies spaced \SI{100}{\mega\hertz} from IFs due to the combs. These are due to nonlinearities in our detection chain, and the high power present in the \SI{100}{\mega\hertz} zeroth order line resulting from the two optical carriers.

To further demonstrate the separability of the two combs, we show proof-of-principle spectroscopy in Fig.~\ref{fig:dualcomb}(d). A fibre Bragg grating (FBG) with centre frequency \SI{191.807}{\tera\hertz} and bandwidth \SI{40}{\giga\hertz} is introduced into the probe arm. Here we arbitrarily choose to use the \SI{7.934}{\giga\hertz} spaced radially polarized comb as the probe comb. The FBG absorbs light from the comb across its bandwidth, resulting in three absent comb lines in the spectrum.

We now examine the relative frequency stability of the dual comb. The two microwave sources providing the comb spacing frequencies, and the \SI{100}{\mega\hertz} r.f.~source driving the AOM are locked to the same \SI{10}{\mega\hertz} clock, allowing accurate assessment of the phase noise due to optical noise in the resonator. This is measured in a similar way as for the single comb, by mixing the beat tone between two comb lines with a local oscillator (from a source that is locked to the same \SI{10}{\mega\hertz} clock as the comb sources). This generates an IF of $\sim \SI{100}{\hertz}$, which is sampled at a frequency of \SI{10}{\kilo\hertz}. The power spectrum of the IF signal is then calculated.

\begin{figure}
\includegraphics{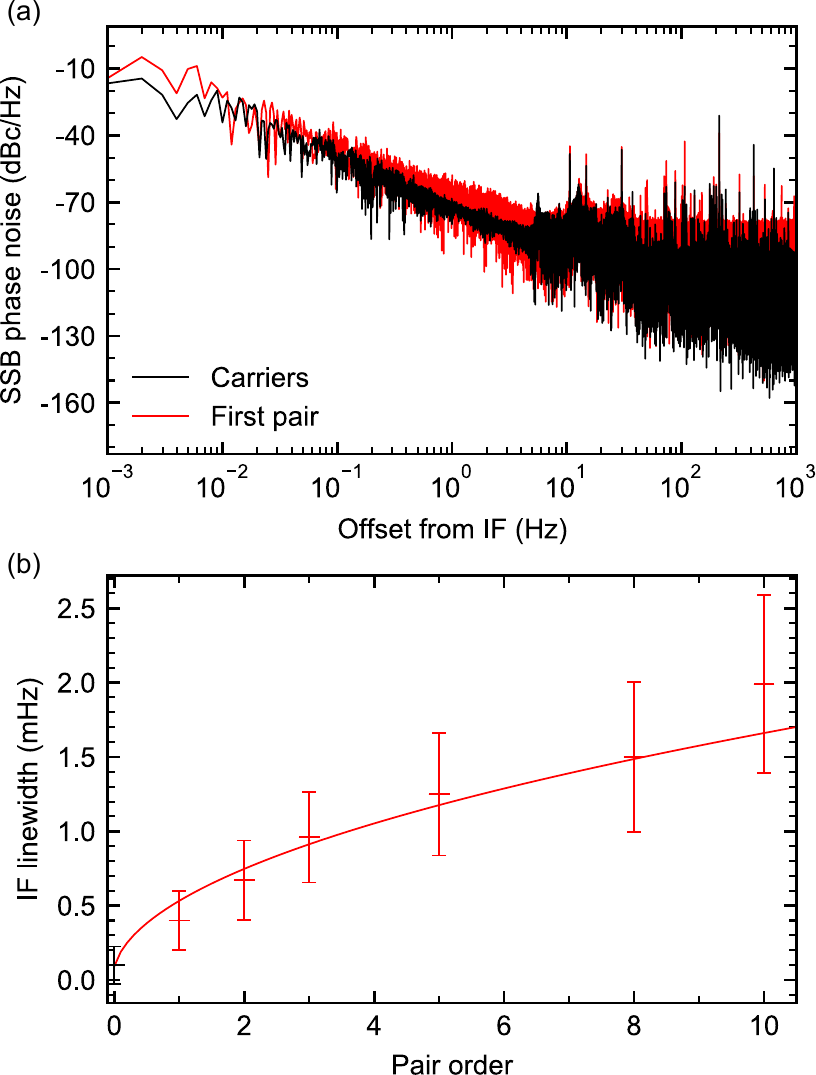}
\caption{\textbf{Relative stability of dual comb.} (a) SSB phase noise of the beat frequencies between the carriers (blue) and first order comb lines (red). (b) IF linewidth vs dual comb line pair order. The fitted curve assumes comb broadening due to incoherent noise in $\chi^{(2)}$.}\label{fig:dualcombstab}
\end{figure}

We measure the linewidths for IF orders between 1 and 10. In Fig.~\ref{fig:dualcombstab}(a) we show the SSB phase noise for the zeroth order IF, resulting from the two optical carriers separated by \SI{100}{\mega\hertz}, and first order IF, generated by comb lines separated by \SI{20}{\mega\hertz}. In Fig.~\ref{fig:dualcombstab}(b) we show linewidth vs generating comb line pair order. We determine an IF linewidth of \SI{0.217(096)}{\milli\hertz} for the zeroth order beat note between the two carriers. This originates from uncorrelated noise on the two fibre arms of the output of the AOM.

Furthermore, we find that the IF linewidth rises with increasing comb line pair order. By assuming that the noise generating processes for each comb are correlated between comblines, but uncorrelated between the two combs and uncorrelated with the AOM, we fit
\begin{align}\label{eqn:dclinewidths}
\sigma_n^2 = \sigma_\textrm{AOM}^2+2n(\sigma_\textrm{mW}^2+\sigma_\textrm{LN}^2)
\end{align}
to the IF linewidths due to the $n$th order pair (see Methods). We find $\sigma_\textrm{LN} = \SI{0.36(08)}{\milli\hertz}$, consistent with the value determined above for the single comb.

The dual comb relative stability we have demonstrated far exceeds the performance of previous unstabilized dual combs~\cite{lucasSpatialMultiplexingSoliton2018}, and also outperforms dual combs stabilized using approaches such as reference lasers stabilized to external cavities~\cite{coddingtonCoherentDualcombSpectroscopy2010}, self referenced combs~\cite{kwonGenerationMultipleUltrastable2020, coddingtonRapidPreciseAbsolute2009}, or combs generated from the same laser cavity~\cite{ideguchiKerrlensModelockedBidirectional2016}. Furthermore, the performance exceeds that obtained by either digital~\cite{trochaUltrafastOpticalRanging2018, zhuTwocolorPhasestableDualcomb2019} or analogue~\cite{ideguchiAdaptiveRealtimeDualcomb2014} post-processing. We ascribe this high frequency stability to the fact that all our generating frequencies are directly derived from phase stable microwave sources, which are all synchronized to the same \SI{10}{\mega\hertz} clock. Comb generation for both combs takes place in the same resonator mode volume, and so noise generated by microwave frequency fluctuations in the \LN{} is common to both combs.

Apart from \LN{}, a wide variety of materials exhibit significant second-order optical nonlinearities. The WGM FSRs are determined by the optical path length around the resonator, which is governed by the resonator diameter and the refractive index for that polarization of light. The repetition rate for the two combs can therefore be chosen by careful selection of resonator material and geometry. Some possible material choices are described in the Supplementary Information. Further tuning would be possible by the application of a d.c.~electric field across the \LN{} ring, either in the axial or polar direction. 

In conclusion, we have demonstrated that the wave-vector dependent nonlinearities present in WGM resonators fabricated from $x$-cut \LN{} can be used to generate frequency combs, which avoids the requirement for a spatially varying microwave field.  They also allow the simultaneous generation of two combs, with different comb line spacings and different polarizations. Because of the spatial multiplexing of the two combs, and the fact that phase locked microwave sources control all relevant frequencies, our combs demonstrate extremely high mutual coherence and are entirely free-running, requiring no special stabilization techniques.  

Moreover, comb generation is deterministic and start-up speed is limited only by the lifetime of the cavity modes. In addition to slow thermal tuning, rapid fine tuning of the centre frequencies and comb spacings could be achieved by application of a d.c.~bias. This device is therefore a platform for simple and cost effective dual comb generation, and represents a step towards deployable ultrastable dual comb-based technologies.

\section{Acknowledgments}

N.J.L. is supported by the MBIE (New Zealand) Endeavour Fund (UOOX1805). We also acknowledge support from the Marsden Fund Grant no.~20-UOO-080. We gratefully acknowledge comments on the manuscript from Dr James Haigh, A.~Prof.~Miro Erkintalo, A.~Prof.~Jevon Longdell and A.~Prof.~Stuart Murdoch.

\section{Competing Interests}

The authors have filed a provisional patent application for aspects of this work at the United States Patent and Trademark Office (application number 63216484). The authors declare no other competing interests.

\section{Data availability}

The data that support the findings of this study are available from the authors on reasonable request.

\section{Methods}

\subsection{Sample design and fabrication}

The microwave cavity was made in two parts; an inner rod and an outer cap. To obtain the correct fundamental mode frequency we based the dimensions on design rules for loop gap cavities (see S.I.), with a gap relative permittivity of $\epsilon_r =57$ representing the $x$-cut \LN{} occupying the electric field mode volume. The initial design was then modelled using COMSOL multiphysics~\cite{comsol}, and design parameters adjusted accordingly.

A ring-shaped \LN{} precursor was cut from a \SI{0.5}{\milli\metre} thick $x$-cut wafer using grinding techniques, and fixed to top of the inner rod using cyanoacrylate. This was cut to shape and size using diamond turning until around \SI{100}{\micro\metre} thick in the radial direction. It was then mechanically polished using \SI{1}{\micro\metre} and \SI{0.25}{\micro\metre} diamond slurry.

Finally the cavity mode frequency and FSRs of the WGM resonator were measured, the outer cap removed, and a small amount of copper in the loop of the loop-gap removed from the fabricated device to achieve the required mode frequency.

\subsection{Linewidth measurement}

To measure the relative stability of frequency comb lines, the spectrum generated by the comb was mixed to r.f.~intermediate frequencies using a Thorlabs DXM20AF fast photodetector with a \SI{20}{\giga\hertz} bandwidth. To select the intermediate frequency of interest, the r.f.~signal was mixed again (Minicircuits ZX05-11X-S+/ZX05-153LH-S+), with the local oscillator (Rohde and Schwarz SMR20) detuned \SI{106.2}{\hertz} from the expected intermediate frequency. The local oscillator linewidth was determined by mixing against an independent microwave source, and found to be narrower than our measurement floor. The resulting \SI{106.2}{\hertz} signal was low pass filtered (\SI{3}{\decibel} cutoff of \SI{15}{\kilo\hertz}) to avoid aliasing then digitized at a sample rate of \SI{10}{\kilo\hertz}. The power spectrum of this signal was then calculated, and the width of the peak at \SI{106.2}{\hertz} determined by fitting a Lorentzian lineshape.

\subsection{Error analysis for fit}

Fitting Eqn \ref{eqn:dclinewidths} to dual comb beat linewidths to determine $\sigma_\textrm{LN}$ requires careful handling of the uncertainties. In particular, $\sigma_\textrm{AOM}$ and $\sigma_\textrm{mW}$ are independently measured parameters in this equation, but also have an uncertainty associated with them. To account for this, we estimate the uncertainty of $\sigma_\textrm{LN}$ by drawing 10000 samples for $\{\sigma_{n}\}$, $\sigma_\textrm{AOM}$ and $\sigma_\textrm{mW}$ from uncorrelated normal distributions having widths equal to the experimentally measured uncertainties, and then fitting for each set to find estimates for $\sigma_\textrm{LN}$. The width of the resulting distribution gives the uncertainty on $\sigma_\textrm{LN}$.

\bibliography{CombLibrary,standards}

\end{document}


\title{Supplementary Information for ``An ultra-stable electro-optic dual frequency comb''}

\author{N. J. Lambert}
\email{nicholas.lambert@otago.ac.nz}
\affiliation{Department of Physics, University of Otago, Dunedin, New Zealand}
\affiliation{The Dodd-Walls Centre for Photonic and Quantum Technologies, New Zealand}
\author{L. S. Trainor}
\affiliation{Department of Physics, University of Otago, Dunedin, New Zealand}
\affiliation{The Dodd-Walls Centre for Photonic and Quantum Technologies, New Zealand}
\author{H. G. L. Schwefel}
\affiliation{Department of Physics, University of Otago, Dunedin, New Zealand}
\affiliation{The Dodd-Walls Centre for Photonic and Quantum Technologies, New Zealand}

\date{\today}

\maketitle
\section{Effective electro-optic nonlinearities}
The nonlinearity experienced by propagating light in a nonlinear medium is dependent on the direction of its wavevector $\mathbf{k}$ in the crystal, and is encapsulated by the electro-optic tensor of the material \cite{boyd2003nonlinear}. However, for whispering gallery modes (WGMs) the direction is not singular. Nevertheless, an effective nonlinearity can be derived for light in modes with a defined polarization and lying in a given crystal plane.

\begin{figure*}[b]
\includegraphics{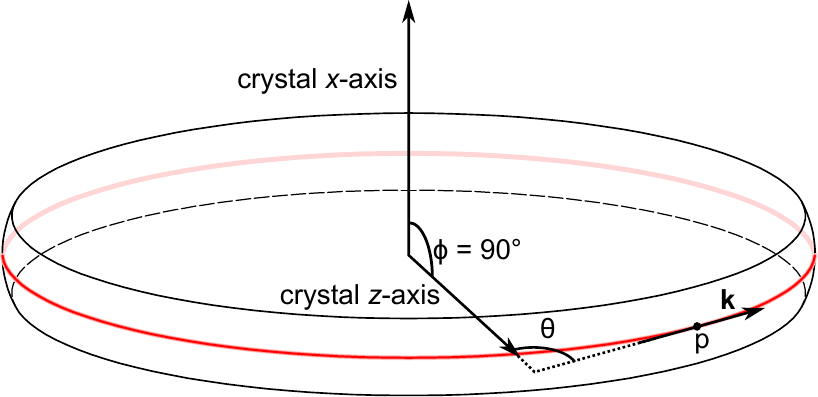}
\caption{Geometry for the calculation of effective nonlinearities. Here we show an $x$-cut resonator. $\theta$ is the angle between the light wavevector $\mathbf{k}$ at point p and the crystal $x$-axis, and $\phi$ is the angle between the wavevector and the $z$-axis. The trajectory of the WGM is shown in red.} \label{fig:geometry}
\end{figure*}

In the following the angle $\theta$ will be the angle the wavevector of the light makes with the optic- or $z$-axis. The angle $\phi$ is then the angle the wavevector of the light makes with the $x$-axis, i.e.\ these are standard Physicist's spherical coordinates (Fig. \ref{fig:geometry}).

We will assume a resonator with the optic axis in the plane of the resonator. The angle $\phi$ is a property of the resonator; it is the angle the $x$-axis makes with the plane of the resonator, i.e.\ we have the following correspondences:
\begin{equation}
\text{$x$-cut} \rightarrow \phi = \SI{90}{\degree},\qquad
\text{$y$-cut} \rightarrow \phi = \SI{0}{\degree}.
\end{equation}
The angle $\theta$ varies as the light propagates around the rim of the resonator. Our whispering-gallery-mode resonator supports two polarizations of modes, which have major field components which are either polarized axially or radially. The two polarizations of light then have the following polarization unit vectors: 
\begin{alignat}{3}
& \text{radial polarization (e)} && \rightarrow \bm{e}_\mathrm{radial} && = (\cos\theta\cos\phi, \cos\theta\sin\phi, -\sin\theta), \\
& \text{axial polarization (o)} && \rightarrow \bm{e}_\mathrm{axial} && = (\sin\phi, -\cos\phi, 0).
\end{alignat}
These are the common extraordinary (e) and ordinary (o) polarization vectors in a uniaxial crystal. 
Furthermore, we see from Fig.~1(b,c) of the main text that the major field component of our microwave resonance is radially polarized.

We will examine the interaction energy between two optical modes and a microwave mode. We will look at the case of upconversion, where one photon in a pump resonance (labelled $p$) and one microwave photon (labelled $\Omega$) are annihilated to create a sideband photon (labelled $+$) with the frequency that is the sum of the two annihilated photons. This interaction energy is proportional to
\begin{equation}
  W \propto \int \chi_{ijk} (E_p)_i (E_+^*)_j (E_\Omega)_k dV, 
\end{equation}
where $(E_a)_j$ is the $j$-component of the positive-frequency electric field of mode $a\in\{p,+,\Omega\}$.
The optical modes are close in frequency and there fit many wavelengths ($\sim 20,000$) around the rim. Furthermore, all of the comb lines of a single comb will be of the same polarization, such that they can be equally spaced a number of free spectral ranges apart. We can therefore approximate $\bm{E}_+ = \bm{E}_p\exp(i [m_+-m_p] \theta)$, where $m_a$ is the number of oscillations mode $a$ makes around the rim of the resonator. That is, the modes look the same in cross section, but the sideband has $(m_+-m_p)$ extra wavelengths around the rim. In our experiment, adjacent comb lines are one free spectral range apart, so $m_+-m_p=1$.

Radially-polarized modes will appear to expand and contract around the resonator's rim due to varying local permittivity.
In addition, the microwave field will have extra variation due to the hole for the coupling prism.
We ignore these effects by assuming that the transverse distribution of the product of the electric fields is rotationally symmetric.
We therefore assume that the angular dependence of our microwave field is $E_\Omega \sim \exp(i m_\Omega \theta)$.
Our interaction energy is then proportional to
\begin{align}
  W & \propto \frac{1}{2\pi}\int_0^{2\pi} \chi_{ijk} (e_p)_i (e_+)_j (e_\Omega)_k \exp(i \theta [m_+-m_p-m_\Omega] ) d\theta
  \\
  \label{eq:chieffdef}
  & \overset{\mathrm{def}}{=} \frac{1}{2\pi}\int_0^{2\pi}\chi(\theta)\exp(i\theta\Delta m) d\theta  \overset{\mathrm{def}}{=} \chi_\mathrm{eff},
\end{align}
where $(e_a)_j$ is the $j$-component of the polarization vector of mode $a$, $\chi(\theta) = \chi_{ijk} (e_p)_i (e_+)_j (e_\Omega)_k$, and $\Delta m=m_+-m_p-m_\Omega$.
$\chi_\mathrm{eff}$ is seen to be the component of the nonlinear susceptibility, which contributes $\Delta m$ momentum quanta.
We now evaluate $\chi(\theta)$ and $\chi_\mathrm{eff}$.

The components of the nonlinear susceptibility in this case are found from the electro-optic tensor, $r$.
The electro-optic tensor is most commonly found in its contracted Voigt form, where the first index refers to the polarization of \emph{both} optical fields at once, using the rule in Table \ref{tab:chi2contractionrule}. The Voigt form is
\begin{table}
\centering
\caption{Voigt-form index-contraction rule}
\begin{tabular}{c || c | c | c | c | c | c }
		$ij$ or $ji$ & $xx$ & $yy$ & $zz$ & $yz$ & $xz$ & $xy$ \\ \hline
		$l$ & 1 & 2 & 3 & 4 & 5 & 6 \\
\end{tabular}
\label{tab:chi2contractionrule}
\end{table}
\begin{equation}
[r_{lk}] = 
\begin{bmatrix}
0&-r_{22}&r_{13}\\
0&r_{22}&r_{13}\\
0&0&r_{33}\\
0&r_{51}&0\\
r_{51}&0&0\\
-r_{22}&0&0
\end{bmatrix}.
\end{equation}
We will refer to $[\chi_{lk}]$, which is the nonlinear susceptibility contracted in an identical way and it has the same structure (zero and nonzero components) as $[r_{lk}]$. These components are associated with nonlinear susceptibilities through the relation
\begin{equation}
  \chi^{(2)}_{ijk} = -\frac{\epsilon_{ii} \epsilon_{jj}}{2\epsilon_0^2}r_{ijk}.
\end{equation}
We therefore have in lithium niobate
\begin{equation}
  \chi_{22} = -\frac{r_{22}}2n_o^4,\quad \chi_{13} = -\frac{r_{13}}2n_o^4,\quad \chi_{33} = -\frac{r_{33}}2n_e^4,\quad \chi_{51} = -\frac{r_{51}}2n_o^2n_e^2.
\end{equation}
Using the electro-optic coefficents from Ref.~\cite{turnerHIGHFREQUENCYELECTRO1966}, and refractive indices from Ref.~\cite{PhysRevB.50.751}, we find that the nonlinear susceptibility has the values in Table \ref{tab:electroopticvalues}.

\begin{table}
\centering
\caption{High-frequency electro-optic coefficient values for LiNbO\textsubscript{3} for optical light at \SI{633}{\nano\meter} and modulation frequency of \SIrange{50}{86}{\mega\hertz} \cite{turnerHIGHFREQUENCYELECTRO1966}. The values are expected to be marginally lower for \SI{1550}{\nano\meter} light \cite{mendez_wavelength_1999}. We then calculate the associated nonlinear susceptibility using these values and the refractive indices at \SI{1550}{\nano\meter} \cite{PhysRevB.50.751}.}
\begin{tabular}{c || c | c | c | c }
		$lk$ & $33$ & $13$ & $51$ & $22$ \\ \hline
		$r_{lk}$ (\si{\pico\meter\per\volt}) & 30.8 & 8.6 & 28 & 3.4 \\ \hline
		$-\chi_{lk}$ (\si{\pico\meter\per\volt}) & 321 & 102 & 312 & 40.4  \\
\end{tabular}
\label{tab:electroopticvalues}
\end{table}

The tensor product of the nonlinear susceptibility with the electric field polarization vectors is then found from
\begin{align}
  \chi(\theta) & = [(e_\mathrm{optical})_l][\chi_{lk}][(e_\Omega)_k],
\end{align}
where the contracted polarization vector for the optical fields is
\begin{equation}
  [(e_\mathrm{optical})_l] = 
    \begin{bmatrix}
    L_{xx}&L_{yy}&L_{zz}&L_{zy}&L_{zx}&L_{xy}
  \end{bmatrix},\quad\text{where}
\end{equation}
\begin{equation}
  L_{ii} = (e_p)_i(e_+)_i,\qquad L_{ij} = (e_p)_i(e_+)_j + (e_p)_j(e_+)_i,\quad i\neq j,
\end{equation}
and the polarization vector for the microwave field is radial.

The two polarization possibilities therefore have a $\chi^{(2)}(\theta)$ that varies with angle as
\begin{align}
\chi_\mathrm{axial}(\theta) = [(e_\mathrm{axial})_l][\chi_{lk}][(e_\mathrm{radial})_k] = &-\chi_{13}\sin\theta + \chi_{22}\cos\theta\sin 3\phi,
\\
\chi_\mathrm{radial}(\theta) = [(e_\mathrm{radial})_l][\chi_{lk}][(e_\mathrm{radial})_k]  = &-\frac{1}{4}\left[3\chi_{33}+2\chi_{51}+\chi_{13}\right]\sin\theta - \frac{3}{4}\chi_{22}\cos\theta\sin 3\phi \\
& + \frac{1}{4}\left[\chi_{33}-2\chi_{51}-\chi_{13}\right]\sin 3\theta - \frac{1}{4}\chi_{22}\cos 3\theta\sin 3\phi. \nonumber
\end{align}

We see here that when the optical field is polarized axially, there are $\sin\theta$ and $\cos\theta$ terms in $\chi(\theta)$. Because $\cos\theta = [\exp(-i\theta)+\exp(i\theta)]/2$, and $\sin\theta = i[\exp(-i\theta)-\exp(i\theta)]/2$,
we see that we must therefore have $\Delta m = \pm 1$ in order to obtain nonzero $\chi_\mathrm{eff}$ from equation \eqref{eq:chieffdef}. In the case of radially-polarized optical fields there are also $\sin 3\theta$ and $\cos 3\theta$ terms, so we could also have $\Delta m = \pm 3$.

Our microwave resonator was designed to support modes with $m_\Omega = 0$ at a frequency of approximately the optical free spectral range. We are therefore looking at the case where $\Delta m = 1$. We can integrate equation \eqref{eq:chieffdef}, and we find for the two optical polarizations
\begin{align}
  \left|\chi_\mathrm{eff,axial}\right| & = \frac{1}{2}\sqrt{\chi_{13}^2 + (\chi_{22}\sin 3\phi)^2}, \\
  \left|\chi_\mathrm{eff,radial}\right| & = \frac{1}{8}\sqrt{(3\chi_{33} + \chi_{13} + 2\chi_{51})^2 + (3\chi_{22}\sin 3\phi)^2}.
\end{align}
For our $x$-cut resonator, we substitute $\phi=\SI{90}{\degree}$ and find $|\chi_\mathrm{eff,axial}|=\SI{55.0}{\pico\meter\per\volt}$ and $|\chi_\mathrm{eff,radial}|=\SI{212}{\pico\meter\per\volt}$.

\section{Toroidal loop-gap resonators}

\begin{figure*}[b]
\includegraphics{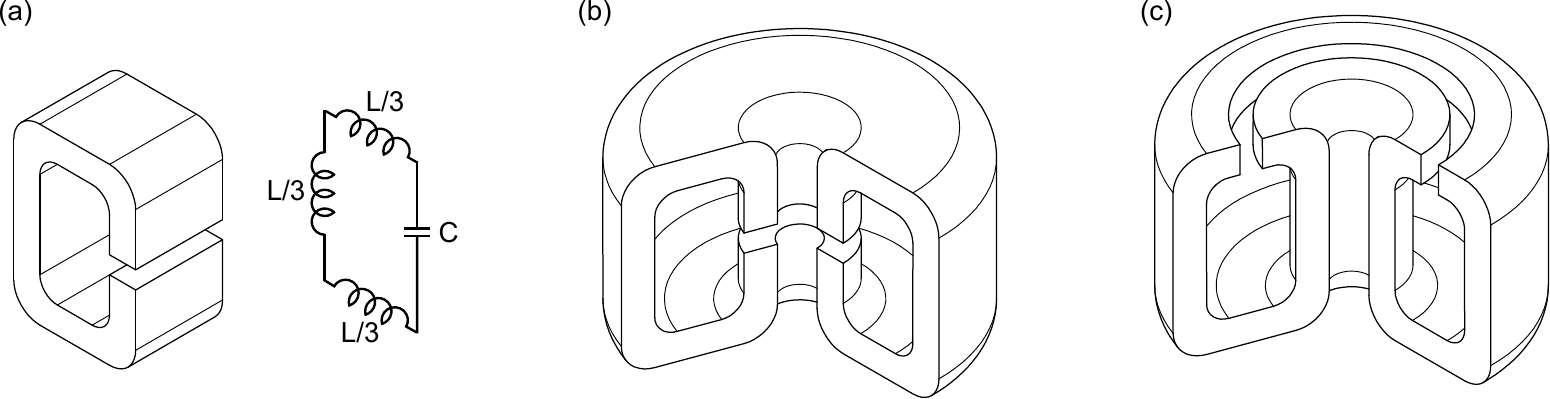}
\caption{Loop gap resonators of various designs. (a) Left, a simple loop gap resonator. Radiative losses from the ends of the resonator can dominate the $Q$-factor. Right, cartoon lumped element model showing the inductance of the loop and capacitance of the gap. (b) By folding the loop gap resonator upon itself such that it becomes a torus, radiative losses can be reduced. (c) Our toroidal loop gap resonator has the gap at the top of the loop, rather than the centre.} \label{fig:loop_gaps}
\end{figure*}

Loop-gap resonators are common resonant elements in high frequency circuits~\cite{pozar2004microwave}. They comprise an inductance, $L$, due to a single loop of current, and a capacitance, $C$ formed from a gap in that loop (Fig.~\ref{fig:loop_gaps}(a)). This constitutes an LC circuit, with resonant frequency $f_r=1/(2\pi \sqrt{LC})$. Resonant modes in loop-gap resonators have the magnetic field component through the loop, and the electric field across the gap.

They typically have moderately high $Q$-factors, with the damping dominated by radiative losses. To reduce these losses, the resonator can be folded into a toroidal shape (Fig.~\ref{fig:loop_gaps}(b)). Our resonator (Fig.~\ref{fig:loop_gaps}(c)) is topologically equivalent to this.

Design equations for loop gap resonators can be found in Ref.~\cite{eatonBiomedicalEPRPart2005}.

\section{Allan variance}

\begin{figure*}
\includegraphics{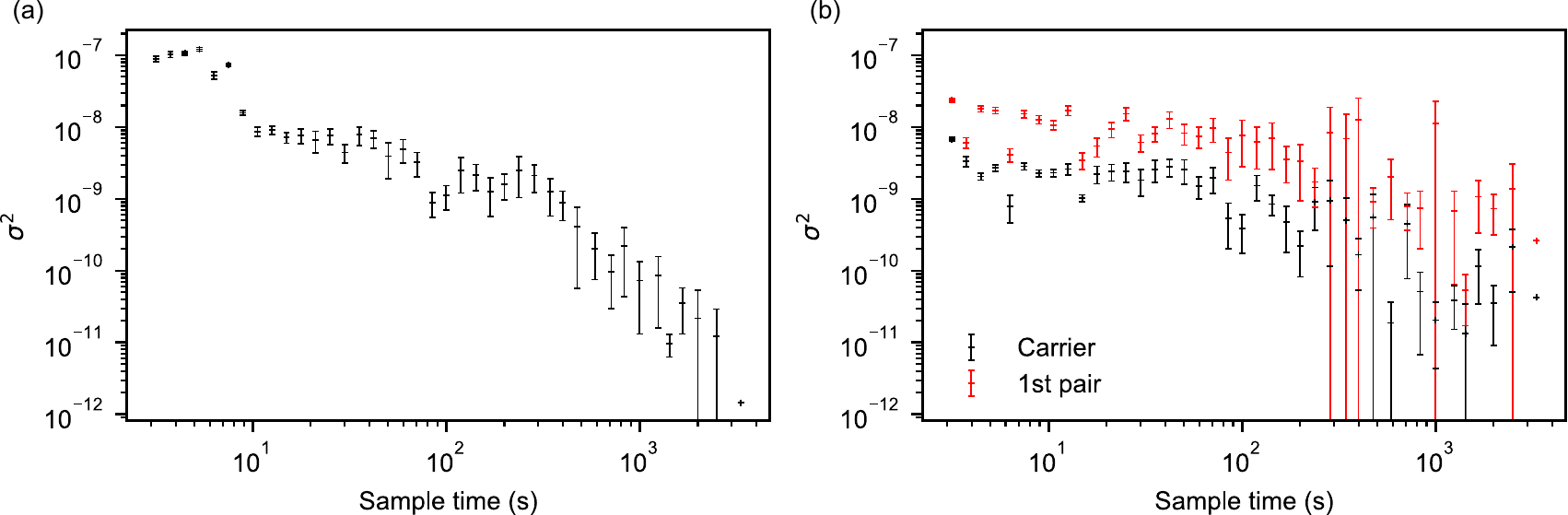}
\caption{Allan variance of comb line beat tones, mixed against a local oscillator to an IF of $\sim\SI{100}{\hertz}$. (a) Allan variance for a single comb. (b) Dual comb Allan variance for the two carriers (black) and the first pair of comb lines (red).} \label{fig:allan}
\end{figure*}

A common metric for the frequency stability of signals is the Allan variance of the signal. In Fig. \ref{fig:allan}
we plot the Allan variance vs sample time for (a) the \SI{7.940}{\giga\hertz} beat note due to the single comb,  (b) the \SI{100}{\mega\hertz} IF from the two carriers (black), and the \SI{15}{\mega\hertz} IF from the first order comb lines (red).

\section{Choice of material and geometry}

Many electro-optic materials can be used in our scheme. In each case one must find which cut of the material will give a good effective nonlinearity for both polarizations. 
The FSRs of the mode families are then determined by the diameter of the microresonator and the birefringence of the material. This gives considerable control over dual comb parameters; we summarise some options in Table \ref{table:materials}.
In each case we have worked out a favourable cut for the material, evaluated the effective nonlinearities at \SI{1550}{\nano\meter} (note that the scheme is relatively wavelength agnostic and other wavelengths could be used), and estimated the percentage difference in FSR between the two mode families. As for lithium niobate, we have used the refractive indices at \SI{1550}{\nano\meter} and the electro-optic coefficient at the wavelengths and frequencies we can find in the literature.
The difference in FSR given is estimated as
\begin{equation}
    \frac{\Delta\mathrm{FSR}}{\mathrm{FSR}} = \frac{\mathrm{FSR}_\mathrm{radial} - \mathrm{FSR}_\mathrm{axial}}{\frac{1}{2}(\mathrm{FSR}_\mathrm{radial}+\mathrm{FSR}_\mathrm{axial})} \approx
    2\frac{\bar{n}_\mathrm{axial}-\bar{n}_\mathrm{radial}}{\bar{n}_\mathrm{axial}+\bar{n}_\mathrm{radial}},
\end{equation}
where $\bar{n}$ is the average bulk refractive index for that polarization. We are therefore neglecting the geometric dispersion (which will be resonator dependent) and the difference between the group and phase refractive indices for this table.

\begin{table}
\caption{Possible materials and resulting comb parameters} \label{table:materials}
\begin{tabular}{lllllll}
\hhline{=======}
Material & point group & cut & $|\chi_\mathrm{eff,radial}|$ (\si{\pico\metre\per\volt})& $|\chi_\mathrm{eff,axial}|$ (\si{\pico\metre\per\volt})& $\Delta$FSR/FSR (\%) & References \\ \hline
Lithium niobate & 3m & $x$-cut & 212 & 55.0 & 1.6 & \cite{turnerHIGHFREQUENCYELECTRO1966,mendez_wavelength_1999,PhysRevB.50.751} \\
Lithium tantalate & 3m & $x$-cut & 164 & 35.2 & -0.090 &  \cite{casson_electro-optic_2004,luther-davies_signs_1970}  \\
Beta barium borate & 3m & $x$-cut& 2.89\footnote{\label{foot:bbo}Assuming all other components are negligible compared to $r_{22}$~\cite{weber_handbook_2002}.} & 3.86\textsuperscript{\ref{foot:bbo}}  & 3.3 & \cite{tamosauskas_transmittance_2018,abarkan_frequency_2003} \\
Potassium titanyl phosphate & mm2 & $y$-cut\footnote{\label{foot:ktp}Biaxial, but local refractive indices when polarized along $x$ and $y$ axes are similar, so coupling should be possible to both polarizations at the same angle. Further, for $y$-cut there is a position on the rim where the effective refractive indices are equal as $n_x<n_y<n_z$.} & 84.7 or 67.8\footnote{\label{foot:ktpsign}$y$-cut: $|2\chi_{51}+\chi_{13}+3\chi_{33}|/8$, larger value if $r_{51}$ is positive. $x$-cut: $|2\chi_{42}+\chi_{23}+3\chi_{33}|/8$, larger value if $r_{42}$ is positive.} & $|\chi_{23}|/2 = 31.3$ & -2.1 & \cite{bierlein_electrooptic_1986,kato_sellmeier_2002} \\
Potassium titanyl phosphate & mm2 & $x$-cut\textsuperscript{\ref{foot:ktp}} & 90.1 or 68.2\textsuperscript{\ref{foot:ktpsign}} & $|\chi_{13}|/2 = 19.6$ & -2.7 & \cite{bierlein_electrooptic_1986,kato_sellmeier_2002} \\
Gallium arsenide & $\bar{4}$3m  &(110) & $3|\chi_{41}|/8 = 36.3$ & $|\chi_{41}|/2 = 48.4$ & 0 & \cite{berseth_electrooptic_1992,skauli_improved_2003} \\ \hhline{=======}
\end{tabular}
\end{table}
\bibliography{CombLibrary}